\documentclass{article}

\usepackage{arxiv}

\usepackage[utf8]{inputenc} % allow utf-8 input
\usepackage[T1]{fontenc}    % use 8-bit T1 fonts
\usepackage{hyperref}       % hyperlinks
\usepackage{url}            % simple URL typesetting
\usepackage{booktabs}       % professional-quality tables
\usepackage{amsfonts}       % blackboard math symbols
\usepackage{nicefrac}       % compact symbols for 1/2, etc.
\usepackage{microtype}      % microtypography
\usepackage{lipsum}
\usepackage{latexsym}
\usepackage{booktabs}
\usepackage{algorithmic}
\usepackage{url}
\usepackage{graphicx}
\usepackage{enumitem}
\usepackage{hyperref}
\usepackage{listings}

\graphicspath{ {images/} }

\title{Which Factors Impact Engagement on News Articles on Facebook?}

\author{
  Marc Faddoul \\
  School of Information, UC Berkeley \\
  {\tt marc.faddoul@berkeley.edu}
}

\begin{document}
\maketitle
\begin{abstract}
Social media is increasingly being used as a news-platform. To reach their intended audience, newspapers need for their articles to be well ranked by Facebook's news-feed algorithm. The number of likes, shares and other reactions determine the lead scoring criteria. This paper will try to assess how the reaction volume is impacted by the following criteria:
\begin{itemize}[leftmargin=*]
    \item Delay between event and post release
    \item Time of the day the post is published
    \item Post format: video, photo or text
\end{itemize}
To isolate the effect of the publication time and post format on a post, we need to control for the news-event and the publishing newspaper. For that end, a news-aggregator is designed and implemented, to group together posts that relate to the same news-event. This tool gave some spin-off results, allowing the ability to map newspapers by similarity and to detect some topic omissions.
\end{abstract}

\section{Introduction}

With two thirds of Americans reading some of their news from social networks ~\cite{Pew}, Facebook is an essentiall distribution channel for newspapers. The news content is in competition to be displayed on a user's news feed. The critical task of content selection and ranking is performed by a private-algorithm; Facebook's feed algorithm. For a post to be widely seen, it has to get a high \emph{relevency score} score from that algorithm. Many parameters are taken into consideration to compute this score; user preferences, time-span since last connection, the device that is being used and even the speed of the internet connection. But one of the main determinants of success is the number of reactions. ~\cite{FYI4} The more likes, shares or other emoji reactions an article gets, the better the relevancy score will be, and the more chances it has to be read by many users. 

This paper will analyze the number of reactions collected by the Facebook publications of the leading English-speaking news platforms (see newspaper list on Appendix table). We will then evaluate how the timing and format of the post impacts the number of reactions. To do so, we need to control the role of two main other factors that impact the reaction volume:

\begin{itemize}[leftmargin=*]
\item \emph{The news-event}: the event that the post is covering.
\item \emph{The news-source}: which newspaper is posting, and how many followers the page has.
\end{itemize}

\section{Dataset}

The posts are collected from 14 of the most followed English-speaking newspapers on Facebook, using the Graph API. All posts published between 2017-11-08 and 2016-06-24 that were still available on the social network in November 2017 are collected. 

The amount of published articles is impressively homogeneous across newspapers, with only \emph{Breitbart} publishing significantly more than the other. The number of followers per page has more variance. (See table in appendix)

For each post, we scrape:

\begin{itemize}[leftmargin=*,noitemsep]
\item The article headline (or video / photo title)
\item The message published with the article 
\item The amount of likes, shares and emoji reactions
\item The date of publication
\end{itemize}

The dataset is be made publicly available on my github, emph{https://github.com/cramdoulfa}. The code of the python scraper will also be published, which allows to easily add news-sources to the analysis.

\section{Impact analysis variables}

\subsection{Dependant Variables}

For this first iteration of the project, we limit our analysis to the reaction volumes. The reaction volume is the sum of the number of likes, shares and other emoji reactions. In a FYI release ~\cite{FYI1}, Facebook said that all reactions will be initially taken into account equally by the feed algorithm. No further public precision on that topic has been made since.

Making the reaction metric more complex, and analysing more specific aspects of the audience reaction is a main direction for further development. Taking into account the number of comments is the reaction volume is a very low hanging fruit. 

Some interesting impact analysis could as well be performed on metrics around sentiment analysis of the comment, that could be supported by the reactions emoji. 

\subsection{Independant Variables}

\subsubsection{The publication timing}

One of the purpose of this research is to analyse the influence of publication time on the virality of an article. Two main aspects will be analysed:
\begin{itemize}[leftmargin=*,noitemsep]
\item \emph{Time in the day:} The time at which the article was published. 
\item \emph{Promptness:} The publication time delay between a given article and the first article published on that same topic (ie from the same cluster). 
\end{itemize}

We can argue that both of these time variables might impact the virality of an article.

There are some prime time ranges throughout the day, where social media activity is enhanced. We test if publishing an article at the \emph{right moment} (see part 5)  has a significant impact on its virality.

Many users follow several newspaper from the media corpus analysed. If a newspaper publishes his article first he might collect the reactions associated to that event, letting behind the articles from the other sources that will loose the 'surprise-effect'. We test this hypothesis by analysing the impact of promptness on virality.

\subsubsection{The post format}

The analysis is original is the sens that it is performed across several post format. These are text articles, videos and photos. Actually, as shown in the appendix table, most newspapers publish a majority of videos on their Facebook page. We test how the format influences the number of reactions.

We infer the format of a post from the category within a facebook page. Post that are qualified as videos can be played directly from Facebook's embedded player. The same is for photos, which can be displayed without leaving the social plateform. What we call text articles are links to the article in the newspaper's website, which may as well include photos and videos. 

\subsection{Control variables}

\subsubsection{The publication source}

Different news sources have different number of followers (see numbers on Appendix table). This has a direct impact on the number of reactions an article can generate. The evolution of the number of followers over time is not available, so the number of followers on 2017-11-08 will be used as an approximation for the whole 17 months period. 

Even controlling for the volume of followers, ~\cite{Bandari} found that the news source is the strongest predictor for the the popularity of an article. This was confirmed by ~\cite{HeadlineMatters}

To control for the impact of the news source, a normalised performance metric is used. For every article, the number of reactions is divided by the average reaction volume of articles from the same source. This score will be called SN-score, for \emph{Source Normalized score}. If an article from CNN has an SN-score of 1.5, it means that it generated 1.5 times more reactions that the average CNN article.

\subsubsection{The news-event}

The news-event is certainly one of the most important criteria when it comes to evaluating an article virality. A cheerful event, such as a national-athlete winning a competition is likely to generate a lot of reactions. Other kinds of news, such as the release of quarterly economic indicator generate less reactions even if they have a thorough coverage from the media corpus.

To control for the excitement potential inherent to each individual news-event, a news aggregator is built. It groups together news article that relate to the same event. This remains an approximation, as every media will relate a given event in a specific way, highlighting or omitting different aspects. Nonetheless, by analysing the performance of different articles within a given cluster, we control for the general topic that is being talked about.

Some parts of the next section are more technical, as they detail the design and implementation of the news aggregator. Is the reader wants to skip it, he can go to section 5 assuming that such a tool was successfully built.

\subsubsection{Limitation: Variables not controlled for}

There are several other factors that can influence the virality, such as wording, illustration, quality of journalistic work, substance depth... All these criteria are highly correlated to the news-source. Newspapers tend to be homogeneous in their style and content quality. Therefor, controlling for the news-source should absorb part of the variance due to the factors listed above.

Meanwhile \cite{HeadlineMatters} analyses the impact of several headline features, such as positivity, magnitude (how extrem the wording is), uniqueness... Including these features in the model would surely improve the results relevance. 

\section{News-Event Aggregation}

\subsection{Task Description}

News aggregation is the task that aims at grouping posts that relate to the same event. This is a fuzzy task, as the scope of an event is never clear.

Some events happen over the scope of several days. For example, at its climax, the Catalan crisis was in the front page of many newspaper during a week. The core of the crisis could be considered as a large event. But every day had a smaller sub-event embedded. For instance, the Spanish government dissolving the parliament or Puigdemont fleeing to Belgium. 

Even when two medias talk about the same specific event, they may relate different aspects of it. 

Therefor, there is not a perfect way to do news-aggregation. The model should allow for some hyper-parameters to adjust for the degree of similarity required for two articles to be classified as the same event.

From a more abstract point of view, news-aggregation is an unsupervised-clustering task. Each post is featurized and represented as a n-dimentional vector. We are trying to cluster this vector dataset into an unknown number of events. The distances between the posts within each cluster have to remain relatively small, as respect to a distance metric that will be defined below.

\subsection{Feature Design}

Here we are looking for the most relevant data representation of a post to perform the clustering task. 

\subsubsection{Time Feature}

The first obvious feature is the time of publication. Posts that deal with the same event are likely to be published within a rather short time-frame, typically a couple of hours or a day.

We will simply use the UNIX timestamp as the feature description of the publication time. (Number of seconds since the 1st of January 1970)

\subsubsection{Text Feature}

The second important aspect is the title (headline for an article, name for a video or a photo) and its attached message. Articles that relate to the same topic should have some words in common. Because we want to process posts from all formats including photos and videos, we need to content ourselves with this short textual description of the event. This is the main difference with other news aggregator found in the literature. They are usually designed for written articles, which allows to process the full article content like in ~\cite{Paraphrase}, ~\cite{Matrix}. ~\cite{HP} even uses further textual enrichment with Wikipedia.

To featurize these textual attributes, we process as follows:

\begin{itemize}[leftmargin=*]
\item \emph{Text normalisation:} Remove punctuation, put all words in lower case. We concatenate the title and the message, as one single text description for each article. \emph{Post description} will refer to this text concatenation.
\item \emph{Vocabulary Generation:} Generate the vocabulary, which is the set of all words that appear at least twice in the dataset. Words that appear only once would not be useful to estimate pairwise distance between articles. 
\item \emph{Term Frequency (TF):} A chunk of text is represented as a integer vector of length equal to the length of the vocabulary (bag of word with one-hot encoding). In our case, the text chunk are the article descriptions. Each dimension corresponds to a word in the dictionary: the value is equal to the number of times that word was found in the description (raw frequency).
\item \emph{Inverse Document Frequency (IDF)}: The inverse document frequency is a word measure of how much information a word provides to discriminate between documents. It is the inverse of the proportion of documents that contain a given word, on a logarithmic scale. It is calculated with the formula: 
\vspace{3mm}
\newline
$\mathrm{IDF}(w, C) =  \log \frac{N}{|\{a \in C: w \in a\}|}$
\vspace{3mm}

Where $w$ is the word, $C$ is the corpus of posts descriptions and N is the total number of posts.
Words with low relevance such as \emph{the, from, have} will have a low IDF whereas words with high relevance such as locations or names will have a high IDF. The IDF vector is unique for the whole corpus.

\item \emph{TF-IDF}: TF-IDF combines the two previous metrics to weight the words found in a description accordingly to their discrimination power. It is obtained by a term-to-term multiplication of the TF vector of each post description with the IDF vector.
\end{itemize}

The resulting TF-IDF vector of the normalized post desciption is the representation of the textual attributes of each post.

Alternatively to this one-hot-vector representation for words, we could have used a word embedding to represent each word. Some word embeddings are designed for the distance between two words in the representation space to be correlated to the semantic distance between two words. This might allow for a better matching of two posts that would describe the same things while using different words. 
In practice, despite is simplicity, TF-IDF seems to be an efficient representation for that task. \cite{Paraphrase}

\subsection{Distance Function}

Most clustering algorithm require a distance metric between data-points. Here the distance between two articles has to take into account both the post description similarity and the time similarity.

Our distance function will be a liner combination of a textal distance and a time distance:

$D(a_{i},  a_{j}) =  D_{t}( a_{i},  a_{j}) +  \delta  D_{w}( a_{i},  a_{j})$

Where $a_{i}$ is an article, and $\delta$ a positive coefficient kept as a model hyper-parameter.

\subsection{Time Distance  $D_{t}$}

The time distance is simply the delay between the publication time of two documents. (A distance of 0.25 corresponds to 6 hours between two publication dates).

Two alternative tweaks have been tested:
\begin{itemize}[leftmargin=*]
    \item \emph{Square the time distance}. Squaring the time delta increases the penalty for bigger time delays between two articles, and diminishes the penalty for shorter delays. The intuition would make sens, but in practice it tends to collapse the clusters a bit too much.
    \item \emph{Normalise the time distance.} In that case, all distances are normalised by the maximum time distance in the dataset. Therefor, all distances are between 0 and 1. 
    Such re-scaling does not impact the clustering task, but it ends up not being very convenient. It results the $\delta$ hyper-parameter setting to be dependant on the time stretch of the dataset. Model tuning ends up being more straight forward without such normalisation.
\end{itemize}

\subsection{Word Distance  $D_{w}$}

The word distance is directly deduced from the TF-IDF representation of the words using the cosine distance. 
The cosine distance is a normalised distance metric that gives the degree of co-linearity between two vectors. With our representation, it measures how similar the vocabulary used in two post descriptions are, taking into account the IDF weighting.
Because it is normalised by the product of the vector norms, the cosine similarity is always a value between -1 and 1. Moreover, our vectors are all positives, which reduces the range to $[0,1]$.

\subsection{Clustering Algorithm}

\subsubsection{Algorithm Choice}

With this metric to estimate the degree of similarity between two articles, we can use an unsupervised clustering algorithm to group the posts together. It is unsupervised as we don't have a sample set of posts for each event we want to create. The most common of these algorithms, such as k-means or Gaussian mixture models require to know the number of clusters to create. In our case, we do not know how many different buckets of posts we want to generate. There are ways to go round this problem by increasing gradually the number of cluster and stop when the cost of having a new cluster over-weights the marginal information gain. (see Information Criteria, ~\cite{Akaike2011}) 

There are also a few clustering algorithms that do not take the number of clusters as parameter. Hierarchical clustering is one of them, where we gradually aggregate the points that are the closest together into clusters, until a condition is reached. Here, the DBSCAN algorithm was used because it allows for a fine tuning of the clusters conditions. The next section details this approach and its motivations.

\subsubsection{DBSCAN algorithm}

\textbf{Principle}
\vspace{2mm}

DBSCAN stands for Density-Based Spatial Clustering of Applications with Noise. ~\cite{DBSCAN} The pseudo code of the core algorithm can be found on \emph{Wikipedia}, but here is the general idea:
\begin{itemize}[leftmargin=*]
    \item DBSCAN groups together the posts that are in the same region of the feature space, with a population density threshold. If a sphere of radius $r$ contains a minimum of $Nmin$ posts, the cloud of points initiates a new cluster.
    \item  Clusters then spread with a density-connectivity criteria: for two points to be part of the same cluster, there must be a chain of points between them. Each of these points must be no further than $r$ from each other, and have at least $Nmin$ points of the cluster in their $r$-neighbourhood. 
    \item $Nmin$ and $r$ are hyper-parameters that control for the posts density within clusters. 
\end{itemize}

\textbf{Motivations}
\vspace{2mm}

The main properties that motivate the choice of this algorithm are the following: 

\begin{itemize}[leftmargin=*]
    \item \emph{Unspecified number of clusters:} As said above, we don't need to specify the number of clusters (ie events).
    \item \emph{Resistance to outliers:} If a point has less than $Nmin$ points in its $r$-neighbourhood, it is considered as noise. Therefor, we don't need to classify every posts. This robustness to outliers is beneficial for our task: some posts talk about specific topic that no other newspaper have addressed, and we don't want these to be assigned to unrealted clusters. 
    \item \emph{Almost deterministic:} As opposed to k-mean or GMM which are very dependant on the initial conditions, DBSCAN is essentially deterministic. Some border points can end up in a different cluster depending on the processing order, but the core of the clusters remain the same.
    \item \emph{Arbitrary distance:} The algorithm can run just with the matrix of pre-computed pairwise distance between points. It is easy to use a custom distance, and it does not have to be a proper distance metric.
    \item  \emph{Arbitrary cluster shape:} A cluster can spread in the time dimension or in the textual dimension, if there is a dense cloud of posts that fills this space. This allows for more flexibility as respect to the cluster shape (useful for a topic that would last longer in time). It is an advantage over k-means, where clusters tend to all be spherical.
\end{itemize}

\textbf{Limitations}
\vspace{2mm}

This last property also implies that with an accurate parameter setting, the algorithm can output long clusters that spread in time around the same topic. Generally speaking, this is a good attribute but it can get problematical for certain edge cases. Sometime, a topic will generate a lot of articles, like during the Las Vegas shooting. Because the topic is dealt with for days, it is aggregated as a big cluster of hundreds of articles over the span of several days. Because it is so densely connected, the periferical areas of the cloud of posts are still above the density threshold. This tends to attract all other posts even if they are far from the core of the cluster, which usually means they are not directly related to the main topic. The bigger a cluster, the easier it is for an post that is little related to the core topic to find some matching neighbours. For instance because Trump has done several declarations about the attack, some articles about another Trump speech the same day may end up on the Las Vegas cluster. To go round that problem, we can make the hyper-parameter more restrictive: make $Nmin$ bigger and $r$ smaller. But these density parameters are shared among all clusters. This would also impact the smaller clusters, thus drastically increasing the number of unclassified posts (outliers). This is the main drawback of this algorithm: \emph{DBSCAN does not allow for big differences in cluster densities.}

This is what motivates the introduction of a customized recursive version of DBSCAN.

\subsubsection{Recursive DBSCAN}

The idea is to re-run DBSCAN on the biggest clusters, to split them further and desagregate distinct news-event that would have been merged.

For that, we need to define what is the limit for a cluster to be considered 'too big', and therefor less likely to be clean (for all posts to be related to the same topic). For that we introduce 2 criteria for a cluster to be oversized:
\begin{itemize}
    \item A maximum number of points per clusters
    \item A maximum width for clusters
\end{itemize}
The width of a cluster is defined as the maximum distance between two of its members.
Here is the pseudo code of the recursive version of DBSCAN:

\begin{lstlisting}[basicstyle=\small]
A = allArticles
while (A not empty):
    A = []
    events = DBSCAN(A, r, N)
    for e in events:
        if (oversize(e) == True):
            for post in e:
                A = A + post
    update(r, N)
    
oversize(e, sizeMax, widthMax):
    if size(e) < sizeMax:
        if width(e) < widthMax:
            return False
    return True

\end{lstlisting}

This new version of the algorithm introduces two more hyper-parameters: the maximum width and the maximum size.

The maximum size is set to 25. Even if some events, such as terror attacks can yield more than 25 posts, this limit will split a big cluster into more specific sub-events: the revelation of the identity of the attacker will for instance be detached as a separate cluster.

The maximum width also acts as a proxy for limiting the time stretch of a cluster. The time distance was defined such that a distance of 1 corresponds to 24h of distance between two articles. We set the maximum width to 1.5, which gives a maximum time stretch of 36 hours. In practice, it will never be that much as it would require the word distance to be 0. 

Eventually, we need to set update rules for $Nmin$ and $r$. $Nmin$ is increased by one at each iteration, and $r$ reduced by 0.05, which allows a slow shrinkage of the oversized clusters until they match the conditions. 

\subsection{Tool performance and limitations}

The hyper-parameters described above are set by hand, by eyeballing the resulting clusters. This is the main limitation of the tool in its current state. A data-set of news-articles aggregated by events ~\cite{UCI} was used validate the algorithmic approach. But because the article types and density are very different in the two data-sets, cross-validating the parameters on the classified data-set to use them in the Facebook corpus did not provide the best results. A test set from the original dataset would be preferred.

Still, with a little parameter-tweaking, this recursive version of the algorithm can cluster a little more than 30\% of the posts, while keeping the clusters almost perfectly clean. Indeed, our analysis does not require the full dataset to be clustered. In any case, a big proportion of posts are unique on their topic and should not be classified. The trade-off between precision and recall was definitely done in favor of the former. In our case, a false negative is just a wasted datapoint, whereas false positives make our analysis noisy.

The litterature introduces more complex approches to event-aggreagtion with self-organizing maps ~\cite{SOM} or Latent Dirichlet allocation ~\cite{Matrix}, which achieve better results. Such improvement would certainly be beneficial to increase the recall and results accuracy. Overall, the news-aggregation performance did not seem to limit the scope of the analysis.

\section{Impact Analysis}

By controlling for the news-source and the news-event we try to assess how the publication time and format impact the virality of an article.

\subsection{Impact of publication time}

We first try to see the impact of publication time on the amount of reactions generated by a post. There are two scales at which this problem can be addressed:
\begin{itemize}[noitemsep]
    \item Time of the day a post has been published
    \item Time a post has been published as respect to the other posts on the same topic
\end{itemize}

\subsubsection*{Impact of the time of the day}

At the scale of a country, the activity on social networks is not steady through the day. The following graph shows the amount of articles published at different times of the day by the American news paper included in the corpus. Times are reported in east-coast time.

We see that publishers concentrate their publications in the evening, when users are most likely to be scrolling on their news feed. What is the impact of adapting the publication time to the users schedule?

We calculate the audience reactivity time-profile of each newspaper: for each news-paper, and at each hour of the day, the average SN-score is computed. (Source-Normalized score, see 1.2).

Now we can evaluate the extent to which news-papers correlate their publication time with their audience reactivity. To do so, we calculate for each source the cosine-similarity between the reactivity time-profile of the audience and the publication time-profile of the newspaper. We call this metric the \emph{audience synchronisation} of a newspaper. 

We then compare these values to the average reaction performance of the newspapers, evaluated as the average number of reactions per articles per 1000 followers. (See Table in Appendix, Synchronisation against Performance)

If we run a linear regression of the performance against the audience synchronisation, we find a extremely significant positive correlation (pvalue of 0.001, slope of 175). An increase of 0.01 in time relevance (difference between USA Today and ABC News) corresponds to an increase of 1.75 more reaction per article, per 1000 followers. This can represent up to a 300\% increase, but this value has to be interpreted carefully. 

First of all, the slope is strongly pushed up by the Daily Mail and Breitbart, two news sources that score great both on performance and time relevance. If we remove these two points, the slope goes down to 91, with a still significant correlation (pvalue = 0.0002).

Second, the correlation does not mean that a better time relevance will cause an increase in performance. It particular, the correlation showed above can be a consequence of the fact that newspapers that pay attention to their timing also endeavor to have attractive headlines or to pick topics that generate more reactions. What this correlation does prove, is that newspaper that perform well tend to have a good timing.

This is an interesting result as timing relevance is a varaible newspapers have good control on. It only requires to monitor the reactivity of the audience throughout the day, and to release the article as close as possible to these prime times. 
Of course this strategy has to be balanced: a newspaper cannot release all of its article at once. With the digitisation of newspaper publication plateforms, article relase has become more and more instantaneous. There sometimes seems to be a race to publish an article as soon as possible when a new piece of information is released. This is what will be explored in the next section: does publishing an article before the other newspapers increases its reactivity?

\subsubsection*{Impact of promptness}

Here, we try to estimate if publishing an article early on when an event happens tends to increase it's virality as compared to the newspaper's standard. 

To do so, we use the article clusters generated with the news-aggregator. For every article, we calculate the delay between its publication time and the first article that was published in its cluster. The first post in a cluster has a delay of zero. Then, we calculate the performance of each article, using the Source-Normalized score to control for the source. With this metric, if a newspaper with high average performance is always the first to publish, it will not bias the results since it's SN-score cannot be always over 1.

When running regressions of delay against SN-score, we find no significant correlation. This is true both if we run the correlations individually on each cluster, or if we aggregate the results. It is true across all formats, and on a specific format. It is also independent of the time frame that we consider: even on the first hour, the 'hot period', publishing an article before the other newspapers does not increases the final number of reactions it will collect, nor does it decreases it.

This non-correlation remains interesting: for a publisher, it does not seem to be worth putting much pressure on the journalists for them to publish instantaneously. It would be interesting to integrate press dispatch to the news aggregator, and see if using this timestamp as the reference for post delays changes this analysis.

\subsection{Impact of post format}

If we look at the posts from the corpus that have generated the biggest reaction volume, we find that they all share a quite straightforward feature.

The top 33 most liked posts from our corpus are videos. And the number 34 is... a photo. The first post that is neither a video nor a video arrives on position 130! It is an article from the Huffington Post about the twitter employee who deleted D. Trump's twitter account.

This is a quite striking result, considering that the corpus is constituted of all the major written press institution in the United States. Every newspaper is concerned by this phenomenon: for all of them, the post most reacted on is a video, except for \emph{Breitbart} whose top-post is a photo.

This shows that for a post to become extremely viral, it has to be visual. But how about the more 'regular' posts about the day-to-day business, that don't pretend to become extremely viral?

Across all articles, video posts get an average SN-score of 1.34, whereas non video have a mere 0.55. Let's try to control by event.

About 70\% of our 1348 news event clusters contained at least one video post. For each of these clusters, we look at the difference between the average SN-scores of the video posts VS the non-video posts. On average, video posts get a 0.644 bigger SN-score than the non-video post on the same topic. That is a very significant difference, that corresponds to a 64\% like increase on an average post.

This analysis stresses that facebook is more adapted to video news than written articles. This phenomenon is already well understood by the newspapers: as one can see on the appendix table, most of them post a majority of videos. 
\vspace{2mm}

\noindent
\emph{Discussion: \textbf{Do videos perform better over written articles due to user preferences or does this result from the design of the feed algorithm?} }
\vspace{2mm}

Videos and photos can be played directly from the player embedded in Facebook. For a written article, it is necessary to follow a link to the publisher's website. Once the user has left the social media site, they may be distracted by other articles and there is no guarantee that they will return to scrolling the social media feed. 

This can be a strong argument for Facebook to push forward videos instead of text articles. When Facebook performs updates on the feed algorithm, they justify it with a short post 'News Feed FYI', but the underlying mechanism remains opaque. A recent update from April 2016 says that posts on which users spend more time will be better ranked. It could be a proxy to promote videos, as they inherently take more time \emph{on the platform} to be consumed. The auto-play feature for videos, which is now the default option on the news feed, is another piece of evidence of Facebook's incentive to promote videos.

To increase visibility, Facebook encourages publishers to "focus on what they do best; making the important and meaningful stories interesting for the audience". ~\cite{Guidelines} Facebook also claims to fight against, 'clickbait headlines that are designed to get attention and lure visitors into clicking on a link' ~\cite{FYI3}. 

Even from the most serious newspapers of the corpus, the most liked articles are videos charged with emotional content but devoid of substance. A school girl singing a song for the Washington Post, \emph{'Monster trucks to the rescue'} for BBC, \emph{'The best hidden tricks for your iPhone'} for the Wall Street Journal... Some are the epitome of clickbait, such as the Dailymail's 5th most liked post which is titled \emph{'HOW do they do this?! :o'} with no further description.

Despite their claim to promote 'original content', and what publishers 'do best' it seems that the platform fails at making any piece of quality journalism viral. The blame cannot be solely put on Facebook, and is certainly also a consequence of the kind of experience that Facebook users are looking for. But the platform's strong promotion of videos, a format \emph{usually} less prone to quality journalism is most likely fostering this phenomenon. Newspapers certainly face a dilemma as to allocating their internal resources accordingly. 

\section{Miscellaneous Results}

\subsection{Topic Omission: a disinformation tool}

The news aggregator is a powerful tool to detect topic omissions. Indeed, if an event cluster contains articles from every news source except for a few, it might be that an important event has not been related by a newspaper.

By looking at the biggest clusters, we can find interesting such cases. 
\begin{itemize}[leftmargin=*,noitemsep]
    \item Some newspapers did not publish articles about the USA's withdrawal of the Paris accord: Fox News, Breitbart, USA Today and the Wall Street Journal.
    \item When Syria signed the agreement, leaving the USA the only country outside the accord, the news was omitted by Fox News, Breitbart and the Wall Street Journal.
    \item When Jeff Session testified before senate, Fox News and Breitbart were the only newspaper not to publish an article on the topic.
    \item When twitter employee temporarily deleted Donald Trump's twitter account, Breitbart was the only American journal not to address the topic.
\end{itemize}

\emph{Note: The newspapers cited above might have still written about these topic on their website without promoting it on facebook. The Daily Mail also omitted all the events listed below, their facebook page happens to be more one of a tabloid than a newspaper.} 

Fake news is legitimately a trending topic these days, and tools are now available to instantly cross-check the reliability of a piece of information. Facebook introduced in October 2017 a new tool to 'provide context about articles' ~\cite{FYI2}. When an article appears on the news feed, an \emph{info} icon allows to retreive information about the news source and to see other articles on the same topic. This tool, that is also built on top of a news aggregation algorithm, is a good step forward against fake news. In practice it is not always available. 

Topic omission is another powerful form of disinformation, which would maybe benefit from more attention. News aggregation could provide interesting feedback about the completeness of one's news provider coverage. Unfortunately, just like for news-checker, this kind of technology would probably benefit most to people who are not most likely to opt in. A tool embedded by default such as the new context tool is probably the only distribution channel that would make such a technology effective.

\subsection{Newspaper Similarity}

The news aggregator provides a way to compare the newspapers in two ways:
\begin{itemize}[leftmargin=*,noitemsep]
    \item The intersection of events addressed by two newspapers
    \item The similarity in the distribution of the 7 different reactions (like, love, haha, sad, angry, wow, share) on a given topic
\end{itemize}

We can featurize a newspaper as a vector of reaction values for each type of reaction and for each cluster. The cosine similarity between these vectors can give insight as to how similarly the audiences of two newspapers react when they read about the same event. 
We now consider each column (newspaper) of the distance matrix as a datapoint, featurized by its distance to every other newspaper. If we perform a Principal Components Analysis (PCA) on these datapoints, and project them on the two main components, we can visalize in two dimensions how the newspapers lay against each other:

\includegraphics[scale=0.55]{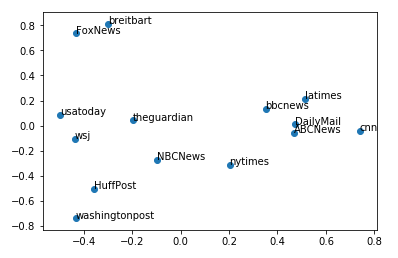}

Such visualisations can provide an interesting way to compare how like-minded the editorial board and the audience of different newspapers are.

\section{Conclusion and Perspectives}

Analysing the factors that impact the reactions volume can provide publishers and account managers with strategical insights to increase their visibility. Our analysis stresses the importance of the publication time and the post format. 

These reactions are one parameter with which Facebook's mysterious algorithm selects the content to which 1.3 billion daily Facebook users will be exposed to. In a very controlled setting, such results could be used to reverse engineer some aspects of the feed algorithm. Reactions and other variables correlated with content visibility can be used as proxies. This analysis hints a strong preference of the platform for videos. Too many variables were left uncontrolled to give any solid conclusion, but further analysis on that direction would be enthralling.

\newpage
\pagebreak

\appendix

\vbox{

        \begin{minipage}[b][0.5\textheight][t]{\textwidth}
        \vspace{0.4in}
\section{Appendix}

\mbox{\textbf{Statistics about the most followed newspaper pages on Facebook in the USA}}

\begin{tabular}{lrrrrrr}
\toprule
{} &  Reactions per &  Followers &  \% of Text &  \% of Video &  \% of Photo & Total \\
{} &  1000 Followers & (Millions) & Posts & Posts & Posts &  Posts \\
Newspaper         &                          &                       &            &              &              &              \\
\midrule
Daily Mail      &                     2.73 &                  13.6 &       41.6 &         58.2 &          0.2 &         1207 \\
Breitbart      &                     2.07 &                   3.5 &       62.7 &         22.7 &         14.6 &         2355 \\
Fox News        &                     1.39 &                  15.6 &       15.7 &         73.6 &         10.7 &         1360 \\
ABC News        &                     0.90 &                  12.1 &       20.4 &         78.1 &          1.5 &         1436 \\
LA times        &                     0.76 &                   2.6 &       66.3 &         24.9 &          8.8 &         1330 \\
NBC News        &                     0.74 &                   9.2 &       26.3 &         72.7 &          1.1 &         1321 \\
The Guardian    &                     0.71 &                   7.6 &       50.1 &         48.2 &          1.7 &         1376 \\
Huffington Post       &                     0.56 &                   9.1 &       32.5 &         66.8 &          0.7 &         1226 \\
Washington Post &                     0.50 &                   6.0 &       39.2 &         60.0 &          0.7 &         1343 \\
USA Today       &                     0.49 &                   8.6 &       26.3 &         69.7 &          4.0 &         1349 \\
CNN           &                     0.38 &                  28.8 &       20.9 &         78.8 &          0.3 &         1458 \\
BBC News        &                     0.35 &                  44.5 &       23.8 &         76.0 &          0.2 &         1402 \\
NY Times        &                     0.34 &                  14.5 &       29.0 &         70.0 &          1.0 &         1480 \\
Wall Street Journal            &                     0.17 &                   5.9 &       78.0 &         19.2 &          2.7 &         1357 \\
\bottomrule
\end{tabular}    

\end{minipage}
    \begin{minipage}[t][0.5\textheight][t]{\textwidth}
    
\mbox{\textbf{Syncronisation with audiance against performance}}

\noindent
\begin{tabular}{lrr}
\toprule
{} & Av Reactions &  Synchronisation  \\
{} & per 1000 foll.    &  with audience\\

\midrule
Daily Mail      &                         2.73 &        0.048003 \\
Breitbart       &                         2.07 &        0.046319 \\
Fox News        &                         1.39 &        0.042774 \\
ABC News        &                         0.90 &        0.045344 \\
LA Times        &                         0.76 &        0.041968 \\
NBC News        &                         0.74 &        0.046489 \\
The Guardian    &                         0.71 &        0.047849 \\
Huff Post       &                         0.56 &        0.044224 \\
Wash. Post      &                         0.50 &        0.043763 \\
USA Today       &                         0.49 &        0.035469 \\
CNN             &                         0.38 &        0.042294 \\
BBC News        &                         0.35 &        0.043253 \\
NY Times        &                         0.34 &        0.041634 \\
WS Journal      &                         0.17 &        0.044584 \\
\bottomrule
\end{tabular}

\vspace{3mm}

\end{minipage}
}
\vspace{3mm}

\clearpage

\section*{Acknowledgments}

This paper was written with the mentorship of David Bamman, as a final project for his course \emph{Natural Language Processing} at the University of California, Berkeley.

\bibliographystyle{unsrt}  
\bibliography{paper}  

\end{document}